# Low-complexity Fusion Filtering for Continuous-Discrete Systems

Seokhyoung Lee and Vladimir Shin

**Abstract**—In this paper, low-complexity distributed fusion filtering algorithm for mixed continuous-discrete multisensory dynamic systems is proposed. To implement the algorithm a new recursive equations for local cross-covariances are derived. To achieve an effective fusion filtering the covariance intersection (CI) algorithm is used. The CI algorithm is useful due to its low-computational complexity for calculation of a big number of cross-covariances between local estimates and matrix weights. Theoretical and numerical examples demonstrate the effectiveness of the covariance intersection algorithm in distributed fusion filtering.

———————————— ◆ ————————————

## 1 INTRODUCTION

RECENTLY a multisensory data fusion has been an interesting topic to increase the accuracy of parameter estimates or estimates of system states. This interest has been motivated by the increased availability of different types of sensors, and as such fusion estimation has the potential for widespread application, since in a range of scenarios, system states or targets are measured by multisensors.

Consequently several distributed fusion architectures and corresponding techniques were presented in [1-4], especially distributed fusion estimation algorithms for finding the best linear combination of the local estimates. The optimal mean-square linear combinations (fusion formulas) of an arbitrary number of local estimates with explicit/implicit expressions for matrix and scalar weights were reported in [4-9]. Furthermore, an effective suboptimal fusion based on the covariance intersection (CI) algorithm was developed in [10, 11].

However, nowadays, with the increasing sophistication of microprocessors, control schemes are being implemented digitally; i.e., in their practical application, received signal are usually measured at a discrete time instants while real signal is continuous. In this case, the above mentioned fusion formulas can not be effectively applicable, because they still require computation of the matrix weights between discrete measurements. In contrast to the optimal fusion formulas, the CI algorithm is more effective and applicable to continuous-discrete system models due to the fact that it computes the weights only at each discrete time instant. To this end, the main purpose of this paper is to compare the CI algorithm with optimal fusion formulas and verify their effectiveness in real implementation.

This paper is organized as follows. In Section 2, optimal fusion formula weighted by matrices is presented. A distributed fusion filtering algorithm for mixed continuous-discrete systems is considered in Section 3. In Section 4, the covariance intersection algorithm is presented. Through a theoretical example, accuracy and low-complexity of the covariance intersection algorithm are verified. In Section 5, a numerical example demonstrates the accuracy and computational efficiency of the covariance intersection algorithm are presented. Finally, a brief conclusion is given in Section 6.

## 2 FUSION FORMULA WITH MATRIX WEIGHTS

Suppose that we have $N$ local estimates of an unknown random signal $x_t \in \mathfrak{R}^n$, i.e.,

$$\hat{x}_t^{(1)},...,\hat{x}_t^{(N)}. \tag{1}$$

Next, let us consider a linear combination of the local estimates with matrix weights. We have

$$\hat{x}_t^{FF} = \sum_{i=1}^{N} C_t^{(i)} \hat{x}_t^{(i)}, \tag{2}$$

where $C_t^{(1)},...,C_t^{(N)}$ are $n \times n$ matrix weights. In addition, from the assumption of an *unbiased* condition, the constraint of the matrix weights is given by

$$\sum_{i=1}^{N} C_t^{(i)} = I_n, \tag{3}$$

where $I_n$ is an $n \times n$ identity matrix. We refer to the fusion formula with matrix weights (2) and (3) as *FF*.

The weights are determined by the mean square criterion [5, 7, 8],

$$J_t = E \left\| x_t - \hat{x}_t^{FF} \right\|^2 \to \min_{C_t^{(i)}}, \tag{4}$$

where the operator $E(\ )$ means expectation. Then the matrix weights $C_t^{(1)},...,C_t^{(N)}$ are given by the following explicit formulas [5]:

————————————————

- *Seokhyoung Lee is with Gwangju Institute of Science and Technology, Gwangju, South Korea.*
- *Vladimir Shin is with Gwangju Institute of Science and Technology, Gwangju, South Korea.*

$$C_t = \left(D_t^T P_{\tilde{x},t}^{-1} D_t\right)^{-1} D_t^T P_{\tilde{x},t}^{-1}, \quad D_t = [I_n \ldots I_n]^T \in \Re^{nN \times n}, \quad (5)$$

where

$$C_t = \begin{bmatrix} C_t^{(1)} & \cdots & C_t^{(N)} \end{bmatrix} \in \Re^{n \times nN}, \quad P_{\tilde{x},t} = \left[P_t^{(ij)}\right] \in \Re^{nN \times nN},$$

$$P_t^{(ij)} = \text{cov}\left(\tilde{x}_t^{(i)}, \tilde{x}_t^{(j)}\right), \quad \tilde{x}_t^{(i)} = x_t - \hat{x}_t^{(i)}, \quad i,j=1,\ldots,N.$$

Furthermore, the fusion error covariance $P_t^{FF} = \text{cov}\left(\tilde{x}_t^{FF}, \tilde{x}_t^{FF}\right)$ is given by

$$P_t^{FF} = \sum_{i,j=1}^{N} C_t^{(i)} P_t^{(ij)} C_t^{(j)T}, \quad \tilde{x}_t^{FF} = x_t - \hat{x}_t^{FF}. \quad (6)$$

As we see in (5), a cross-covariance $P_t^{(ij)}$ is an important factor to implement FF. However, note that the determination of $P_t^{(ij)}$ depends on a concrete model of the random signal $x_t$. So, in this paper, our interest is a cross-covariance $P_t^{(ij)}$ for a random signal x in dynamic systems. In next section, we deal with a stochastic multisensory dynamic system model in detail as an application of (2) and (5).

## 3 FUSION FILTERING IN CONTINUOUS SYSTEMS ON DISCRETE MEASUREMENTS

Let us consider a linear continuous-time stochastic system with N sensors. These sensors measure at discrete times $t_1, t_2, \ldots$. The state-space model of the system takes the form:

$$\begin{aligned}
\dot{x}_t &= F_t x_t + G_t w_t, \quad x_t \in \Re^n, \\
y_{t_k}^{(i)} &= H^{(i)} x_{t_k} + v_{t_k}^{(i)}, \quad y_{t_k}^{(i)} \in \Re^{m_i}, \quad i=1,\ldots,N, \\
k &= 0,1,\ldots,
\end{aligned} \quad (7)$$

where $w_t$ is a zero-mean white Gaussian system noise with intensity $Q_t$, and $v_{t_k}^{(i)}, i=1,\ldots,N$ are zero-mean uncorrelated white Gaussian sequences, i.e., $v_{t_k}^{(i)} \sim N\left(0, R_{t_k}^{(i)}\right), i=1,\ldots,N$, and the initial condition $x_0 \sim N\left(\bar{x}_0, P_0\right)$.

For individual (local) sensors $y_{t_k}^{(i)}$ the system (7) can be divided into N subsystems with common state $x_t$. Each subsystem is described as

$$\begin{aligned}
\dot{x}_t &= F_t x_t + G_t w_t, \\
y_{t_k}^{(i)} &= H_{t_k}^{(i)} x_{t_k} + v_{t_k}^{(i)},
\end{aligned} \quad (8)$$

where the index $i$ is fixed.

Then using the subsystem (8), the local filtering estimate $\hat{x}_{t_k}^{(i)} = \hat{x}_{t_k|t_k}^{(i)}$ of the state $x_{t_k}$ and corresponding error covariance $P_{t_k}^{(ii)} = P_{t_k|t_k}^{(ii)}$ are described by continuous-discrete Kalman filter equations [12, 13]:

Measurement-Update Equations at $t=t_k$:

$$\begin{aligned}
\hat{x}_{t_k|t_k}^{(i)} &= \hat{x}_{t_k|t_{k-1}}^{(i)} + K_{t_k}^{(i)}\left(y_{t_k}^{(i)} - H_{t_k}^{(i)} \hat{x}_{t_k|t_{k-1}}^{(i)}\right), \quad \hat{x}_{0|0}^{(i)} = \bar{x}_0, \\
P_{t_k|t_k}^{(ii)} &= \left(I_n - K_{t_k}^{(i)} H_{t_k}^{(i)}\right) P_{t_k|t_{k-1}}^{(ii)}, \\
K_{t_k}^{(i)} &= P_{t_k|t_{k-1}}^{(ii)} H_{t_k}^{(i)T}\left[H_{t_k}^{(i)} P_{t_k|t_{k-1}}^{(ii)} H_{t_k}^{(i)T} + R_{t_k}^{(i)}\right]^{-1}, \quad P_{0|0}^{(ii)} = P_0,
\end{aligned} \quad (9)$$

where $K_{t_k}^{(i)}$ is a local Kalman filter gain, and the time updates $\hat{x}_{t_k|t_{k-1}}^{(i)}, P_{t_k|t_{k-1}}^{(ii)}$ are obtained by the following differential equations:

Time-Update Equations Between Measurements:

$$\begin{aligned}
\dot{\hat{x}}_{t|t_{k-1}}^{(i)} &= F_t \hat{x}_{t|t_{k-1}}^{(i)}, \quad \hat{x}_{t_{k-1}|t_{k-1}}^{(i)} = \hat{x}_{t_{k-1}}^{(i)}, \quad t \in [t_{k-1}, t_k], \\
\dot{P}_{t|t_{k-1}}^{(ii)} &= F_t P_{t|t_{k-1}}^{(ii)} + P_{t|t_{k-1}}^{(ii)} F_t^T + G_t Q_t G_t^T, \quad P_{t_{k-1}|t_{k-1}}^{(ii)} = P_{t_{k-1}}^{(ii)}.
\end{aligned} \quad (10)$$

Using (9) and (10) we have N local estimates $\hat{x}_{t_k}^{(1)}, \ldots, \hat{x}_{t_k}^{(N)}$ and the corresponding local error-covariances $P_{t_k}^{(11)}, \ldots, P_{t_k}^{(NN)}$. Then the fused estimate $\hat{x}_{t_k}^{FF}$ at $t=t_k$ is calculated by using FF:

$$\hat{x}_{t_k}^{FF} = \sum_{i=1}^{N} C_{t_k}^{(i)} \hat{x}_{t_k}^{(i)}. \quad (11)$$

To compute the matrix weights $C_{t_k}^{(i)}$ in (5) and (11), the exact local cross-covariances $P_{t_k}^{(ij)}, i,j=1,\ldots,N; i \neq j$ can be obtained by solving the following measurement- and time-update equations [5, 8]:

< Measurement − Update Equation >

$$\begin{aligned}
P_{t_k|t_k}^{(ij)} &= \left(I_n - K_{t_k}^{(i)} H^{(i)}\right) P_{t_k|t_{k-1}}^{(ij)} \left(I_n - K_{t_k}^{(j)} H^{(j)}\right)^T, \\
P_{t_k|t_k}^{(ij)} &= P_{t_k}^{(ij)}, \quad P_0^{(ij)} = P_0,
\end{aligned} \quad (12)$$

< Time − Update Equation >

$$\dot{P}_{t|t_{k-1}}^{(ij)} = F_t P_{t|t_{k-1}}^{(ij)} + P_{t|t_{k-1}}^{(ij)} F_t^T + G_t Q_t G_t^T, \quad t \in [t_{k-1}, t_k],$$

$$i,j=1,\ldots,N, \quad i \neq j.$$

Then, the distributed fusion filtering (11) can be implemented using (5), (9), (10), and (12). However in a computational complexity point of view, the above fusion algorithm (11), (12) is not effective, because the solution of the matrix differential equations (10), (12) for all $i,j=1,\ldots,N$, especially in the large number of the sensors, $N \gg 1$, takes a lot of time. Therefore, we focus on how to simplify the calculation of the equation (12).

In the next section, we introduce the covariance intersection algorithm, which does not consider the cross-covariances $P_t^{(ij)}, i \neq j$.

## 4 FUSION FILTERING USING COVARIANCE INTERSECTION ALGORITHM

### 4.1 Covariance Intersection Algorithm

To achieve low-complexity fusion, the covariance intersection (CI) algorithm was proposed for fusion estimation. It takes the same form as the weighted sum, i.e.,

$$\hat{x}_{t_k}^{CI} = \sum_{i=1}^{N} W_{t_k}^{(i)} \hat{x}_{t_k}^{(i)}, \quad W_{t_k}^{(i)} = M_{t_k} \cdot \omega_{t_k}^{(i)} P_{t_k}^{(ii)^{-1}},$$

$$M_{t_k} = \left(\sum_{i=1}^{N} \omega_{t_k}^{(i)} P_{t_k}^{(ii)^{-1}}\right)^{-1}, \quad \omega_{t_k}^{(i)} = \det\left(P_{t_k}^{(ii)^{-1}}\right) / \sum_{j=1}^{N} \det\left(P_{t_k}^{(jj)^{-1}}\right). \quad (13)$$

From (13), only the local error-covariances $P_{t_k}^{(ii)}, i=1,\ldots,N$ are used to calculate the matrix weights

$W_{t_k}^{(i)}$, i=1,...,N for fusion. Therefore, as will be shown, the CI algorithm has low estimation accuracy, but low computational complexity.

### 4.2 Efficiency of Covariance Intersection Algorithm

Since the CI algorithm is a robust fusion algorithm and provides a bound on the estimation accuracy, sometimes the accuracy of fusion estimation can be low. However, basically, the CI algorithm is low computational complexity because cross-covariances between two local estimates are ignored.

The following example makes it possible to compare the specific accuracies and effectiveness between FF and the CI algorithm.

*Example: Fusion Estimation in a Steady-state Regime*

Let us consider a scalar system with two sensors. Using (7) the system is given by

$$\dot{x}_t = -x_t + w_t, \quad w_t \sim (0, q), \quad t \geq 0,$$
$$y_{t_k}^{(i)} = x_{t_k} + v_{t_k}^{(i)}, \quad v_{t_k}^{(i)} \sim N(0, r^{(i)}), \quad k=0,1,..., \quad i=1,2. \quad (14)$$

Then, local error-covariances $P_\infty^{(11)}$, $P_\infty^{(22)}$ and the cross-covariance $P_\infty^{(12)}$ in steady-state regime are obtained [12, 13]:

$$P_\infty^{(ii)} = \frac{qr^{(i)}}{q+2r^{(i)}}, \quad P_\infty^{(12)} = \frac{2r^{(1)}r^{(2)}q}{(q+2r^{(1)})(q+2r^{(2)})}, \quad i=1,2. \quad (15)$$

Using (15), we can obtain specific corresponding steady-state values for fusion weights $C_\infty^{(i)}$, $W_\infty^{(i)}$, i=1,2:

$$C_\infty^{(1)} = \frac{P_\infty^{(22)} - P_\infty^{(12)}}{P_\infty^{(11)} + P_\infty^{(22)} - 2P_\infty^{(12)}}, \quad C_\infty^{(2)} = \frac{P_\infty^{(11)} - P_\infty^{(12)}}{P_\infty^{(11)} + P_\infty^{(22)} - 2P_\infty^{(12)}},$$
$$W_\infty^{(1)} = \frac{(P_\infty^{(22)})^2}{(P_\infty^{(11)})^2 + (P_\infty^{(22)})^2}, \quad W_\infty^{(2)} = \frac{(P_\infty^{(11)})^2}{(P_\infty^{(11)})^2 + (P_\infty^{(22)})^2}. \quad (16)$$

Let us assume that q=1, $r^{(1)}=5$, $r^{(2)}=2$, and use (6). Then we can take specific fusion errors:

$$P_\infty^{FF} = 0.3896, \quad P_\infty^{CI} = 0.3925. \quad (17)$$

In (17), we see that the difference between the fusion error-covariances of FF and CI algorithm is about 0.7%. This means, the accuracy of the CI algorithm is high for this example. Furthermore, as shown in (15) and (16), the calculation of $P_\infty^{(12)}$ is not necessary for the weights $W_\infty^{(i)}$, i=1,2, and thus the CI algorithm can have low computational complexity. Therefore, the CI algorithm is considered more effective for implementation.

## 5 NUMERICAL EXAMPLE

In this section we compare the performance of two fusion filters based on FF and CI algorithm through a numerical example.

Let us consider the longitudinal dynamics of an aircraft. This can be approximately represented by the harmonic oscillator system:

$$\dot{x}_t = \begin{bmatrix} 0 & 1 \\ -\omega_n^2 & -2\xi\omega_n \end{bmatrix} x_t + \begin{bmatrix} 0 \\ 1 \end{bmatrix} w_t, \quad (18)$$

where $x_t = \begin{bmatrix} \theta_t & \dot{\theta}_t \end{bmatrix}^T$ and $\theta_t$ is pitch angle. $\omega_n$ and $\xi$ are natural oscillating frequency and damping ratio, respectively. The process noise $w_t$ represents wind gusts which influence pitch rate.

In addition, three identical sensors (one of them is a main sensor and the others are reserved) measuring at discrete times $t_k$ with different measurement errors are given:

$$y_{t_k}^{(i)} = [1 \quad 0] x_{t_k} + v_{t_k}^{(i)}, i=1,...,N=3,$$
$$k=0,1,...,k_T=50, \quad (19)$$

where $v_{t_k}^{(i)}$, i=1,2,3 are scalar white Gaussian sequences (sensor errors) with variances $r^{(i)}$, i=1,2,3, i.e., $r^{(1)}=1$, $r^{(2)}=2$, $r^{(3)}=3$.

Let us assume that $w_t$ is a zero-mean continuous white Gaussian process (system noise) with intensity q = 2. In addition, we utilize the forth-order Runge-Kutta method to solve differential equations (10) and (12) between measurements with an integration unit time increment $\Delta t = 0.01$.

Figure 1 illustrates the mean square errors (MSE) of the fusion estimate $\hat{\theta}_{t_k}$ based on FF and CI, i.e.,

$$P_{t_k}^{FF} = E(\theta_{t_k} - \hat{\theta}_{t_k}^{FF})^2, \quad P_{t_k}^{CI} = E(\theta_{t_k} - \hat{\theta}_{t_k}^{CI})^2. \quad (20)$$

Moreover Table 1 shows implemented CPU times of two discussed algorithms using a computer with the following specifications: Intel® Pentium® 4, CPU 3 GHz, 1GB RAM.

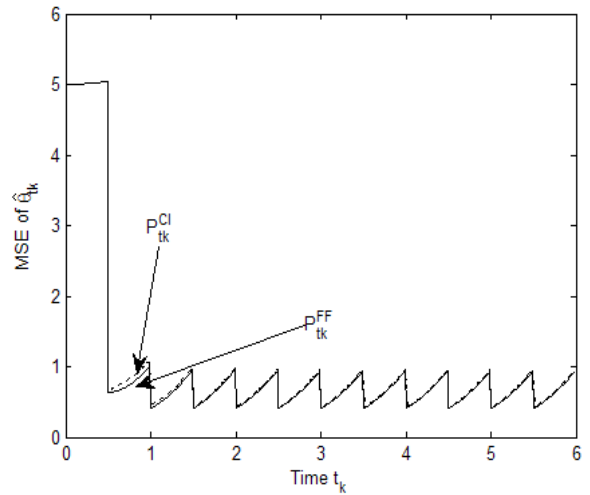

Fig. 1. MSEs of the fusion estimate $\hat{\theta}_{t_k}$ using FF and CI

As shown in Figure 1, we observe that MSEs corresponding to the CI algorithm and FF are very close, and this point was already verified from the theoretical example in Subsection 4.2. Thus, we can say that the CI algorithm is accurate for FF. Furthermore the CI algorithm is almost 64 times faster than FF in performing the fusion filtering as shown in Table 1. Therefore, we see that the CI algorithm is a useful algorithm for real-time implementation, especially when applications are modeled by continuous-discrete time systems.

## 6 CONCLUSION

We presented distributed fusion filtering algorithm for continuous-time dynamic model. This algorithm is based on the optimal fusion formula (FF) and suboptimal CI algorithm. Specific performances of FFs and CI were verified using a theoretical example. Accuracies of the proposed distributed fusion filtering using the FF and CI are very closely. However, the CI algorithm is more effective than the FF with respect to computational complexity, especially on discrete measurements for continuous-time signals. Numerical example demonstrated the effectiveness of the CI algorithm.

Generally when the FF is applied, it takes time progressively longer to fuse the local estimates as the number of sensors increases. In contrast to the use of FF, the weights of the CI only depend on local error covariances, and thus the CI algorithm has lower computational complexity, and can be more easily implemented on discrete measurements in real-time applications.

TABLE 1
IMPLEMENTATION TIMES OF FUSION FILTERS

| Fusion Formulas for Fusion Filtering | CPU Times (s) |
|---|---|
| FF | 0.2115 |
| CI | 0.0033 |

## REFERENCES

[1] Y. Bar-Shalom, and X. R. Li, *Multi-target Multi-sensor Tracking: Principles and Techniques*, Storrs: YBS Publishing, 1995.

[2] M. E. Liggins, C. Y. Chong, I. Kadar *et al.*, "Distributed fusion architecture and algorithms for target tracking," *Proceedings of IEEE.* pp. 95-107, 1997.

[3] Y. M. Zhu, *Multisensor Decision and Estimation Fusion*, Boston: Kluwer, 2002.

[4] X. R. Li, Y. M. Zhu, J. Wang *et al.*, "Optimal Linear Estimation Fusion - Part I: Unified Fusion Rules," *IEEE Transactions on Information Theory*, vol. 49, no. 9, pp. 2192-2208, 2003.

[5] J. Zhou, Y. Zhu, Z. You *et al.*, "An Efficient Algorithm for Optimal Linear Estimation Fusion in Distributed Multisensor Systems," *IEEE Transactions on Systems, Man, and Cybernetics*, vol. 36, no. 5, pp. 1000-1009, 2006.

[6] Z. Deng, Y. Gao, L. Mao *et al.*, "New approach to information fusion steady-state Kalman filtering," *Automatica*, vol. 41, no. 10, pp. 1695-1707, 2005.

[7] Y. Bar-Shalom, and L. Campo, "The Effect of the Common Process Noise on the Two-sensor Fused-track Covariance," *IEEE Transactions on Aerospace and Electronic Systems*, vol. 22, no. 6, pp. 803-805, 1986.

[8] V. I. Shin, Y. Lee, and T. S. Choi, "Generalized Millman's Formula and its Applications for Estimation Problems," *Signal Processing*, vol. 86, no. 2, pp. 257-266, 2006.

[9] V. Shin, G. Shevlyakov, and K. Kim, "A New Fusion Formula and Its Application to Continuous-Time Linear Systems with Multisensor Environment," *Computational Statistics and Data Analysis*, vol. 52, no. 2, pp. 840-854, 2007.

[10] C. Y. Chong, and S. Mori, "Convex Combination and Covariance Intersection Algorithms in Distributed Fusion," *Proceedings of the 4th International Conference of Information Fusion.* pp. WeA22:1-8, 2001.

[11] D. Franken, and A. Hupper, "Improved fast covariance intersection for distributed data fusion," *8th International Conference on Information Fusion.* pp. WbA23:1-7, 2005.

[12] F. L. Lewis, *Optimal Estimation with an Introduction to Stochastic Control Theory*, New York: Wiley & Sons, 1986.

[13] A. H. Jazwinski, *Stochastic Processes and Filtering Theory*, New York: Academic Press, 1970.

**Seokhyoung Lee** received the B.S. degree in Electrical and Electronics Engineering from Kyungpook National University, Korea in 2006. He received the M.S. degree in Mechatronics from Gwangju Institute of Science and Technology, Korea, in 2007. He is currently working toward a Ph.D. degree in the Department of Information and Mechatronics, Gwangju Institute of Science and Technology, Korea. His research interests are estimation, filtering, data fusion, and stochastic system control.

**Vladimir Shin** received the B.Sc. degree and M.Sc. degree in applied mathematics from Moscow State Aviation Institute, Russia, in 1977 and 1979, respectively. In 1985 he received his Ph.D. degree in mathematics at the Institute of Control Science, Russian Academy of Sciences, Moscow. From 1984 to 1999 he was Head of the Statistical Methods Lab. at the Institute of Informatics Problems, Russian Academy of Sciences, Moscow. During 1995 he was Visiting Professor in the Department of Mathematics at Korea Advanced Institute of Science and Technology. From 1999 to 2002 he was Member of Technical Staff at the Samsung Institute of Technology. He is currently an Associate Professor at Gwangju Institute of Science and Technology, South Korea. His research interests include estimation, filtering, tracking, data fusion, stochastic control, identification, and other multidimensional data processing. He has authored or coauthored over 80 papers in these fields.